\documentclass[prl,twocolumn,showpacs,aps]{revtex4-1}
\usepackage{graphicx}
\usepackage{dcolumn}
\usepackage{bm}
\usepackage{amsmath}
\usepackage{amssymb}
\usepackage[usenames,dvipsnames]{xcolor}
\usepackage{color}

\begin{document}


\title{Topological Nonlinear Anomalous Nernst Effect in Strained Transition Metal Dichalcogenides}
\author{Xiao-Qin Yu$^{1}$}
\email{yuxiaoqin@hnu.edu.cn}

\author{Zhen-Gang Zhu$^{2,3,7}$}
\email{zgzhu@ucas.ac.cn}

\author{Jhih-Shih You$^{4}$}
\email{jhihshihyou@gmail.com}

\author{Tony Low$^{5}$}
\email{tlow@umn.edu}

\author{Gang Su$^{3,6,7}$}
\email{gsu@ucas.ac.cn}

\affiliation{$^{1}$ School of Physics and Electronics, Hunan University, Changsha 410082, China.\\
$^{2}$School of Electronic, Electrical and Communication Engineering, University of Chinese Academy of
Sciences, Beijing 100049, China. \\
$^{3}$ Theoretical Condensed Matter Physics and Computational Materials Physics Laboratory, College of Physical Sciences, University of Chinese Academy of Sciences, Beijing 100049, China.\\
$^{4}$ Institute for Theoretical Solid State Physics, IFW Dresden, Helmholtzstr. 20, 01069 Dresden, Germany.\\
$^{5}$ Department of Electrical and Computer Engineering, University of Minnesota, Minneapolis, Minnesota 55455, USA.\\
$^{6}$ Kavli Institute of Theoretical Sciences, University of Chinese Academy of Sciences, Beijing 100049, China.\\
$^{7}$ CAS Center for Excellence in Topological Quantum Computation, University of Chinese Academy of Sciences, Beijing 100190, China.}

\begin{abstract}
We theoretically analyze the non-linear anomalous Nernst effect as the second-order response of temperature gradient by using the semiclassical framework of electron dynamics. We find that a non-linear current can be generated transverse to the applied temperature gradient in time-reversal-symmetry materials with broken inversion symmetry. This effect  has a quantum
origin arising from the Berry curvature of states near the Fermi surface. We discuss the non-linear Nernst effect in transition metal dichalcogenides~(TMDCs) under the application of uniaxial strain. In particular, we predict that under fixed chemical potential in TMDCs, the non-linear Nernst current exhibits a transition from $T^{-2}$ temperature dependence in low temperature regime to a linear $T$-dependence in high temperature.
\end{abstract}

\maketitle
Modern condensed matter physics looks for new phenomena that arise from
the properties of wavefunctions beyond the band structure of materials. A remarkable example of such phenomena
is provided by the local~(geometrical) properties of wavefunctions, defined as the Berry curvature~\cite{Xiao2010,Nagaosa2010}.
The Berry curvature has profound effects on thermoelectric transport by statistical force~(the gradient of temperature), e.g. anomalous Nernst effect~\cite{Xiao2006, Zhang2008,Zhang2009,Zhu2013}, which describes the generation of a charge current in the transverse direction to an applied temperature gradient in the longitudinal direction for a system with broken time-reversal symmetry. Similar to the Hall conductivity~\cite{Deyo2009,Moore2010,Low2015,Sodemann2015}, the Nernst coefficient of a time-reversal-symmetric material vanishes in the first-order response to temperature gradient.
However, the nonlinear responses could manifest distinctive behaviors and have become promising tools for understanding novel materials with
low crystalline symmetry in experiment~\cite{Eginligil2015,Quereda2018,Xu2018,Ma2018,Kang2018}.
Especially, it has recently been shown that nonlinear Hall current as a second-order response to electric field can occur in a wide class of time-reversal invariant and inversion breaking materials~\cite{Deyo2009,Moore2010,Low2015,Sodemann2015}.
The Boltzmann approach has been used to describe the nonlinear transverse current in terms
of the Berry curvature dipole~(BCD), a first-order moment of the Berry curvature over the occupied states in momentum space~\cite{Low2015,Sodemann2015,Nakai2018,Nandy2019}.

%

Monolayer MoS$_2$ and other transition-metal dichalcogenides~(TMDCs) represent a new class of two-dimensional~(2D) materials, intrinsically behaving as semiconductors. Due to lack of an inversion center, their nonvanishing Berry curvature in each valley and strong spin-orbit coupling~(SOC) lead to a series of valley and spin related anomalous transport phenomena, such as valley~(spin) Hall  effect~\cite{Wang2012,Xiao2012,Mak2014} and valley~(spin) Nernst effect~\cite{Yu2015,Yu2017}. However, $H$-structure monolayer TMDCs cannot host linear and nonlinear currents flowing transverse to the driving forces~(electric field or temperature gradient) due to the presence of time reversal and $C_{3v}$ symmetry~\cite{Sodemann2015,You2018}.
When uniaxial strain~\cite{Sodemann2015,You2018,Lee2017,QingyuZhang2013,RodriguezLopez2018,Fang2018} is applied, this $C_{3v}$ symmetry can be reduced to only a single mirror symmetry, in which case the nonlinear Hall current induced by BCD is allowed~\cite{Sodemann2015,You2018}.
In two dimensional materials, it has been shown that the largest symmetry for a nonvanishing BCD is a single mirror symmetry line~\cite{Sodemann2015}.

Recently, it has been reported that intrinsic anomalous non-linear thermoelectric effect can emerge due to orbital troidal moment which breaks time-reversal~(T) and inversion~(I), but retains their combined symmetry~\cite{Gao2018}. However, it is natural to ask whether the nonlinear Nernst current could occur as a second-order response to temperature gradient in novel materials with  time reversal symmetry, but with broken inversion symmetry.
In this paper, we will demonstrate that the nonlinear Nernst effect is determined by a pseudotensorial quantity that, in a similar way as the BCD~\cite{Low2015,Sodemann2015}, is also related to the Berry curvature~(BC) near the Fermi surface  to be discussed below. Our study shows that the thermally driven non-linear current can be generated  perpendicular to applied temperature gradient in the uniaxial strained 1$H$-TMDCs. The proposed nonlinear effect here would also be expected to occur in other materials whose inversion symmetry is broken but with time reversal symmetry, such as topological insulators and three-dimensional Weyl semimetals~\cite{Low2015,Sodemann2015,Morimoto2016,deJuan2017,Tsirkin2018,Zhang2018,Zhang2018_1,You2018,Facio2018,Du2018,Du2018_1,Shi2019}.

In the following, our study builds upon a seminal work by Xiao \textit{et .al.}~\cite{Xiao2006}, which shows that statistical force, arising from the temperature gradient $\nabla T$, gives rise to a correction to the orbital magnetization and an anomalous charge current $\mathbf{j}_\text{A}$ perpendicular to the temperature gradient
\begin{eqnarray}
\mathbf{j}_\text{A} &=&-\frac{\nabla T}{T}\times \frac{e}{\hbar}\int [d\mathbf{k}] \mathbf{\Omega}\left(\mathbf{k}\right)\left[\left(E_\mathbf{k}-\mu_e\right)f\left(\mathbf{k}\right)\right.  \notag\\
&+& \left.k_\text{B}T \ln\left(1+e^{-\beta\left(E_\mathbf{k}-\mu_e\right)}\right)\right],
\label{j-1}
\end{eqnarray}
where $\beta=1/k_\text{B}T$, $k_\text{B}$ represents the Boltzmann constant, $E_\mathbf{k}$ is the band energy, $\mu_e$ indicates  the chemical potential,  $\hbar$ is the Planck constant, $\nabla T$ denotes the temperature gradient, $\mathbf{\Omega}\left(\mathbf{k}\right)$ is the Berry curvature, 
and $\int [d\mathbf{k}]$ is the shorthand for $\int d\mathbf{k}/(2 \pi)^{d}$.
Here, we have generalized the electron distribution in the formalism of thermally induced anomalous charge current $\mathbf{j}_\text{A}$ in Ref.~\cite{Xiao2006} into the non-equilibrium one, namely replacing equilibrium Fermi-Dirac distribution function $f_{0}\left(\mathbf{k}\right)=1\left/\left(\exp{\frac{E_\mathbf{k}-\mu_e}{k_\text{B}T}}+1\right)\right.$ by non-equilibrium distribution function $f\left(\mathbf{k}\right)$ (details can be found {\color{red}{in Ref.\cite{SI}}}).
In fact, the  current $\mathbf{j}_\text{A}$ is zero in the linear response regime, i.e. $f=f_{0}$, for a time-reversal-invariant system, owing to the relation  $\mathbf{\Omega}\left(\mathbf{k}\right)=-\mathbf{\Omega}\left(-\mathbf{k}\right)$ guaranteed by time-reversal symmetry and the equal occupation for states at $\mathbf{k}$ and $-\mathbf{k}$ in linear regime.
However, when the distribution is non-equilibrium, the transverse anomalous charge current $\mathbf{j}_\text{A}$ can survive as the second-order response to the temperature gradient, which we will show below.

With the relaxation time approximation, the Boltzmann equation for the distribution of electrons in absence of electric field is~\cite{mahan_many_2000}
\begin{equation}
f-f_{0}=-\tau \frac{\partial f}{\partial r_a}\cdot v_{a},
\label{Bol-eq3}
\end{equation}
where $\tau$ represents the relaxation time, $v_{a}$ and $r_{a}$ denote the $a$ component of the velocity and coordinate position of electrons, respectively. We are interested in the response up to second order in temperature gradient, and hence have the nonequilibrium distribution function $f\approx f _{0}+f_{1}+f_{2}$ with the term $f_{n}$ understood to vanish as $(\partial{T}/\partial \mathbf{r}_{a})^{n}$. After a series of careful derivation~(see details in {\color{red}{Ref.\cite{SI}}}), one obtains
\begin{equation}
\begin{aligned}
f_{1}&=\frac{\tau}{T\hbar}\left(E_\mathbf{k}-\mu_e\right)\frac{\partial f_{0}}{\partial k_{a}} \partial_{a} T,\\
f_{2}&=-\frac{\tau^{2}}{T\hbar}\left(E_\mathbf{k}-\mu_e\right)v_{b}\frac{\partial f_{0}}{\partial k_{a}}\left(\partial_{ab}T-\frac{1}{T}\partial_{a}T\partial_{b}T\right)\\
&+\frac{\tau^{2}}{\hbar^{2} T^{2}}\left(E_{k}-\mu_e\right)^{2}\frac{\partial^{2} f_{0}}{\partial k_{a}\partial k_{b}}\partial_{a}T\partial_{b}T,\\
\label{f1}
\end{aligned}
\end{equation}
where $\partial_{a}={\partial}/{\partial r_{a}}$ and $\partial_{ab}={\partial^{2}}/{\partial r_{a}\partial r_{b}}$. Based on Eqs. (\ref{j-1}) and (\ref{f1}), the nonlinear Nernst-like current density $\mathbf{j}^\text{nl}_\text{A}$ (where the subscript ``{A}"/ superscript ``{nl}" refer to anomalous /nonlinear, respectively) in the $a$ direction, as the response to second order in temperature gradient, is found to be
\begin{equation}
\begin{aligned}
\left[\textbf{j}^\text{nl}_\text{A}\right]_{a}&=-\varepsilon_{abc}\frac{\partial_{b}T\partial_{d}T}{T}\frac{\tau e}{T\hbar^{2}}\int[d \mathbf{k}]\left(E_\mathbf{k}-\mu_e\right)^{2}\frac{\partial f_{0}}{\partial k_{d}}\Omega_{c},\\
\end{aligned}
\label{current-dip}
\end{equation}
where $\varepsilon_{abc}$ is Levi-Civita symbol. Eq. (\ref{current-dip}) indicates that the current $\left[\textbf{j}^\text{nl}_\text{A}\right]_{a}$ is linearly proportional to the relaxation time $\tau$  and  a pseudotensorial quantity, 
defined as
\begin{equation}
\begin{aligned}
 \Lambda_{cd}^\text{T}&=-\frac{1}{T^{2}}\int[d\mathbf{k}]\left(E_\mathbf{k}-\mu_e\right)^{2}\frac{\partial{f_{0}}\left(\mathbf{k}\right)}
{\partial k_{d}}\Omega_{c}.\\
\end{aligned}
\label{N}
\end{equation}
The presence of the factor $\partial_{k_{d}}{f_{0}}$ implies that the non-linear current is associated with a "Fermi-surface" contribution, because only the states close to the Fermi surface contribute to the integral in the low temperature. The novel pseudotensorial quantity $\Lambda_{cd}^\text{T}$  features an extra factor ${\left(E_\mathbf{k}-\mu_e\right)^{2}}/{T^{2}}$ in the integral, therefore it has different physical meaning in contrast to  the Berry curvature dipole, $D_{cd}=\int [d\mathbf{k}]f_{0}\partial_{k_{d}} \Omega_{c}=-\int [d\mathbf{k}] \Omega_{c}\partial_{k_{d}} f_{0}$ \cite{Sodemann2015} although they play a similar role in generating current density (see Eq. (\ref{j-2D1})) as that in Ref. \onlinecite{Sodemann2015}. 


In two dimensional materials, the Berry curvature $\mathbf{\Omega}$  is reduced from a pseudovector to a pseudoscalar and only the component perpendicular to
the plane, namely $\Omega_{c=z}$, can be nonzero. Hence, pseudotensorial quantity $\Lambda^\text{T}_{cd}$ behaves as a pseudo-vector contained in the two-dimensional plane:
\begin{equation}
\begin{aligned}
\Lambda^\text{T}_{d}
&=-\frac{1}{T^2}\int [d \mathbf{k}]\frac{\partial f_{0}\left(\mathbf{k}\right)}{\partial k_{d}}\left(E_\mathbf{k}-\mu_e\right)^2\Omega_{z}\left(\mathbf{k}\right),
\end{aligned}
\label{D-e-1}
\end{equation}
where $d=x$, or $y$ indicating the component of $\mathbf{\Lambda}^\text{T}$ in $d$ direction.
The $\mathbf{\Lambda}^\text{T}$ is tied to the underlying crystal symmetries. The largest symmetry of a 2D crystal that allows for nonvanishing $\mathbf{\Lambda}^\text{T}$ is a single mirror line~(i.e., a mirror plane that is orthogonal to the 2D crystal)~\cite{Sodemann2015}. The presence of mirror symmetry accompanying with time reversal symmetry would force the $\Lambda^\text{T}_{d}$ to be orthogonal to the mirror plane, which can be easily illustrated by  analyzing the symmetric/ antisymmetric properties for $k_{d}\rightarrow -k_{d}$ of the energy dispersion $E(k_{d})$ ($E(\mathbf{k})=E_\mathbf{k}$) and Berry curvature $\Omega_{z}\left(k_{d}\right)$, respectively.

Let us consider a time-reversal invariant system in which only mirror symmetry $M_{d }$ exists. Time-reversal symmetry requires that the energy dispersion respects $E\left(\mathbf{k}\right)=E\left(-\mathbf{k}\right)$ and the  mirror symmetry $M_{d}$ imposes the constraint $E\left(k_{d},k_{d_{\perp}}\right)=E\left(-k_{d},k_{d_{\perp}}\right)$, where $d_{\perp}$ indicates the coordinate axis orthogonal to axis-$d$ in 2D plane.  Both constraints on the energy dispersion also imply the relation $E\left(k_{d},k_{d_{\perp}}\right)=E\left(k_{d},-k_{d_\perp}\right)$. Therefore, it is evident that the partial derivative of Fermi-Dirac distribution function $g(\mathbf{k})={\partial f_{0}\left(\mathbf{k}\right)}/{\partial k_{d \, \text{or} \, d_{\perp}}}$ is an odd function with respect to $k_{d}$ and $k_{d_{\perp}}$, namely $g(k_{d})=-g(-k_{d})$ and $g(k_{d_{\perp}})=-g(-k_{d_{\perp}})$. Meanwhile, the mirror symmetry $M_{d}$ requires
 $\Omega_{z}\left(k_{d},k_{d_{\perp}}\right)=-\Omega_{z}\left(-k_{d},k_{d_{\perp}}\right)$,
and time reversal symmetry leads to $\Omega_{z}(-\mathbf{k})=-\Omega_{z}(\mathbf{k})$. When combining the mirror symmetry and the time reversal symmetry, we get  $\Omega_{z}\left(k_{d},k_{d_{\perp}}\right)=\Omega_{z}\left(k_{d},-k_{d_{\perp}}\right)$. Hence, the integrand in Eq.(\ref{D-e-1}) is even  for $\Lambda^\text{T}_{d}$ but odd for $\Lambda^\text{T}_{d_{\perp}}$, indicating that only the $d$ component of $\mathbf{\Lambda}^\text{T}$ could be nonzero.

In fact, the current can be expressed in vector notation
as
\begin{equation}
\textbf{j}^\text{nl}_\text{A}=\frac{e\tau}{\hbar^{2}}(z\times\nabla T)\left(\nabla T\cdot \mathbf{\Lambda}^\text{T}\right).
\label{j-2D1}
\end{equation}
When only a single mirror symmetry exists, the linear thermally driven transport coefficient tensor has its principal axis aligned with the mirror line and $\mathbf{\Lambda}^\text{T}$ is forced to be orthogonal to mirror line. Consequently, According to Eq. (\ref{j-2D1}), when the temperature gradient is perpendicular to the mirror line, one can recognize that the current that flows perpendicular to temperature gradient has a quantum origin arising from $\mathbf{\Lambda}^\text{T}.$ 

One of candidate 2D materials to observe the nonlinear Nernst effect is monolayer transition-metal dichalcogenides. The $C_{3v}$ symmetry of these crystals with $1H$ structure would force the pseudo-vector quantity  $\mathbf{\Lambda}^\text{T}$ to vanish.  However,  application of uniaxial strain can reduce this symmetry and leave only a single mirror operation, in which case the proposed effect would be observed.

\begin{figure}
\centering
\includegraphics[width=1.0\linewidth,clip]{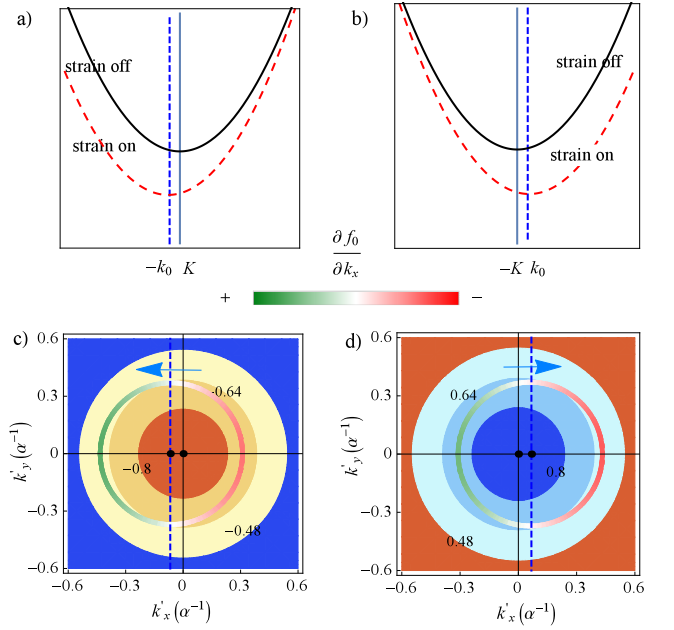}
\caption{(a)(b) Schematic of energy dispersion of a uniaxially deformed TMDCs. (c)(d) Berry curvature $\Omega_{\tau_{v} c}\left(\mathbf{k}\right)$ and the strain-influenced Fermi-surface (solid circle) for the conduction band at the $K$ (-$K$) valley, respectively. The blue arrow denotes the strained induced shift of the band extrema with respect to $\mathbf{k}^{\prime}=0$. The colour scale for the solid circle show schematics of ${\partial f_{0}}/{\partial k_{x}}$ distribution at the Fermi energy. $\alpha$ is the lattice constant.}
\label{SD}
\end{figure}

Under uniform uniaxial strain along high-symmetry, the strained-dependent Hamiltonian of TMDCs around K or -K valley is~\cite{Lee2017,Rostami2015}
\begin{equation}
\hat{H}=\frac{\Delta^{\prime}}{2}\hat{\sigma}_{z}+v_{F}\hbar\left(\tau_{v} k^{\prime}_{x}\hat{\sigma}_{x}+k^{\prime}_{y}\hat{\sigma}_{y}\right)+\tau_{v}\gamma v_{F}\hbar\left(u_{xx}-u_{yy}\right)k^{\prime}_{x},
\label{Hamil-1}
\end{equation}
where $\tau_{v}(=\pm 1)$ is the valley index, $\hat{\sigma}$ denote the Pauli matrices for the two basis functions of the energy band, $v_{F}$ represents the Fermi velocity and $\Delta^{\prime}$ is the strain-modified energy band gap. Due to the strain-induced fictitious vector potential $\mathbf{A}\propto  (u_{yy}-u_{xx},0),$ we substitute the canonical momentum $\mathbf{k}^{\prime}=\mathbf{k}-e \tau_{v} \mathbf{A}$ for the momentum $\mathbf{k}$ which is measured relatively to the valley K ($\tau_{v}=1$) or valley -K ($\tau_{v}=-1$) point of the Brillouin zone.
 The third term that does not couple to the sublattice pseudospin is strain-dependent with $\gamma$ and $u_{ij}$ denoting, respectively, a dimensionless parameter and a strain tensor element. The SOC has been ignored since we consider only n-type
TMDCs and the SOC is weak in the lowest-energy conduction bands of TMDCs~\cite{Liu2013}.  The energy eigenvalues are
\begin{equation}
\begin{aligned}
E_{\tau_{v} n}\left(\mathbf{k}^{\prime}\right)&=\tau_{v}\gamma v_{F}\hbar\left(u_{xx}-u_{yy}\right) k^{\prime}_{x}+n \epsilon_{1}\left(k^{\prime}\right),\\
\epsilon_{1}\left(k^{\prime}\right)&=\sqrt{\left(\frac{\Delta^{\prime}}{2}\right)^{2}+\left(v_{F}\hbar k^{\prime}\right)^{2}},
\end{aligned}
\label{Eigen}
\end{equation}
where $n(=\pm 1)$ is the band index. The Berry curvature of band is determined by $\Omega_{\tau_{v} n}\left(\mathbf{k}\right)=\hat{\mathbf{z}}\cdot \nabla_\mathbf{k}\times \langle u_{\tau_{v} n}|i\nabla_\mathbf{k}|u_{\tau_{v} n} \rangle$ for 2D materials, where $u_{\tau_{v} n}$ is the eigenfunction
of the Hamiltonian, 
For the massive Dirac fermions described by the effective Hamiltonian in Eq. (\ref{Hamil-1}), the Berry curvature of the conduction band is
\begin{equation}
\Omega_{\tau_{v},n=1}\left(\mathbf{k}^{\prime}\right)=-\tau_{v} \frac{v^{2}_{F}\hbar^{2}\Delta^{\prime}}{4\left[\epsilon_{1}\left(k^{\prime}\right)\right]^{3}}.
\label{cond}
\end{equation}
In the valence band, we have $\Omega_{\tau_{v}, n=-1}\left(\mathbf{k}^{\prime}\right)=-\Omega_{\tau_{v} ,n=1}\left(\mathbf{k}^{\prime}\right)$.
For relatively small strain level, the second line of Eq. (\ref{Eigen}) becomes $\epsilon_{1}\left(k^{\prime}\right)\approx \frac{\Delta^{\prime}}{2}+\frac{(\hbar k^{\prime})^{2}}{2m}$, where  $m=\frac{\Delta ^{\prime}}{2v_{F}^{2}}$ is band mass. With this approximation the energy dispersion can be expressed as $
E_{\tau_{v},n=1}\left(\mathbf{k}^{\prime}\right)\approx \frac{\Delta^{\prime}}{2}+\frac{\hbar^{2}\left(\mathbf{k}^{\prime}+\tau_{v} \mathbf{k}_{0}\right)^{2}}{2m},$
where the shift of the band extrema with respect to $\mathbf{k}^{\prime}$ is determined by $\mathbf{k}_{0}=k_{0}\hat{\mathbf{x}}=\frac{\gamma\left(u_{xx}-u_{yy}\right)\Delta^{\prime}}{2\hbar v_{F}}\hat{\mathbf{x}}$.


Building on the above analytical expressions for energy dispersion and Berry curvature, we present an analysis for $\mathbf{\Lambda}^\text{T}$ that describes the nonlinear anomalous Nernst effect. The quantity $\mathbf{\Lambda}^\text{T}$ for TMDCs in conduction band is given as
\begin{equation}
\begin{aligned}
{\Lambda}^\text{T}_{d}=-\frac{2}{T^2}\sum_{\tau_v,n=1}\int [d \mathbf{k}^{\prime}]\left[E_{\tau_{v}n}\left({\mathbf{k}^\prime}\right)-\mu_{e}\right]^2\frac{\partial f_{0,\tau_{v}}}{\partial {{k}^{\prime}_{d}}}
\Omega_{\tau_{v} n}\left(\mathbf{k}^{\prime}\right),\\
\end{aligned}
\label{D}
\end{equation}
where 2 is for spin. It should be noted that ${\Lambda}^\text{T}_{d}$ becomes zero when $T$ approaches zero since ${\partial f_{0,\tau_{v}}}/{\partial {{k}^{\prime}_{d}}}\rightarrow- \delta\left(E_{\tau_{v}n}\left({\mathbf{k}^\prime}\right)-\mu_{e}\right) \frac{\partial E_{\tau_{v}n}\left({\mathbf{k}^\prime}\right)}\partial {{k}^{\prime}_{d}}$. In the following, we will only consider the nonlinear Nernst effect at finite temperature. Application of the uniaxial tensile strain along the zigzag direction ($x$ direction)  indicates that only the $x$ component of $\mathbf{\Lambda}^\text{T}$ in Eq. (\ref{D}) could be nonzero due to the mirror symmetry $M_{x}$. To better understand the physics, let us first look at the case of one valley. For each valley, ${\partial f_{0,\tau_{v}}}/{\partial k^{\prime}_{x}}=-[(1-f_{0,\tau_{v}})/f_{0,\tau_{v}}]\partial E_{\tau_{v},n=1}/{\partial k^{\prime}_{x}}$ is equal but opposite on both sides of the Fermi surface~[Fig. \ref{SD} (c)(d)]. The Berry curvature of the conduction band [Eq. (\ref{cond})] is also isotropic for each valley in $\mathbf{k}^{\prime}$.  If there is no uniaxial strain, this circulating pattern resulting from the BC and ${\partial f_{0,\tau_{v}}}/{\partial k^{\prime}_{x}}$  will have exact cancellation around Fermi energy, giving rise to a vanishing $\mathbf{\Lambda}^\text{T}$ at each valley. Under the uniaxial strain along $x$ direction, however, the band extrema are shifted from the original $K(-K)$ point to opposite directions in $k_{x}$-axis~[Fig. \ref{SD} (a)(b)].
As a result, the center of Fermi sphere for each valley will no longer coincide with the extrema of the Berry curvature~[Fig. \ref{SD} (c)(d)] in $k_{x}$ direction, leading to the non-zero $\Lambda^\text{T}_{x}$ for single valley.

As illustrated in Fig. \ref{SD}, it is visible that $g_{\tau_{v}}\left(\mathbf{k}\right)={\partial f_{0,\tau_{v}}}/{\partial k_{x}}$ has opposite sign on the opposite side of Fermi energy for different valleys, namely $g_{K}\left(k_{x},k_{y}\right)=-g_{-K}\left(-k_{x},k_{y}\right)$, which is entailed by the mirror symmetry. Furthermore due to time-reversal symmetry, the Berry curvature at the two valleys takes opposite values. Therefore, $\Lambda^\text{T}_{x}$ at two valleys can have same sign and contribute additively,
resulting in the total ${\Lambda}^\text{T}_{x}$ as
\begin{equation}
\begin{aligned}
\Lambda^{\text{T}}_{x}=-\zeta u_{xx}\left[\left(\frac{k_\text{B}T}{\Delta^{\prime}}Q_{1}\left(\beta\mu\right)+Q_{2}\left(\beta\mu\right)\right)\right],
\label{Lam_end}
\end{aligned}
\end{equation}
with
\begin{equation}
\begin{aligned}
Q_{1}\left(\beta\mu\right)&=2\beta^{2}\mu^{2}\ln\left(1+e^{\beta\mu}\right)+4\mu \beta\text{Li}_{2}\left(-e^{\beta\mu}\right)\\
&-4\text{Li}_{3}\left[-e^{\beta\mu}\right],\\
Q_{2}\left(\beta\mu\right)&=\frac{1}{3}\pi^{2}-2\text{Li}_{2}\left[(1+e^{\beta\mu})^{-1}\right]
-\left[\ln\left(1+e^{\beta\mu}\right)\right]^{2},
\end{aligned}
\end{equation}
where 
$\beta=1/{k_{B}T}$, $\text{Li}_{n}(x)=\sum_{k=1}^{\infty}\frac{x^{k}}{k^{n}}$ is the polylogarithm. And the chemical potential $\mu=\mu_e-{\Delta^{\prime}}/{2}$ is measured from the bottom of the conduction band. The presence of $\Lambda^\text{T}_{x}$ indicates that with $C_{3v}$ symmetry broken by uniaxial strain, a net transverse current occurs as the second-order response to the longitudinal temperature gradient in TMDCs. Besides, this net transverse current is dominant in this time reversal-invariant system, since there is no transverse current in the linear response regime.
  The typical scale $\zeta=6k_\text{B}^{2} v_{F}\hbar\gamma /(\Delta^{\prime})$ characterizing $\Lambda^{T}_{x}$ is 3.838 $k^2_\text{B}$${\AA }$ for MoS$_2$, where we use a Fermi velocity $v_{F}\approx6.559\times 10^{5}$m$/$s, $\gamma=0.2708$ and the strain modified energy $\Delta^{\prime}\approx 1.82 eV$ at relatively small strain levels~\cite{Rostami2015}.

\begin{figure}
\centering
\includegraphics[width=1.0\linewidth,clip]{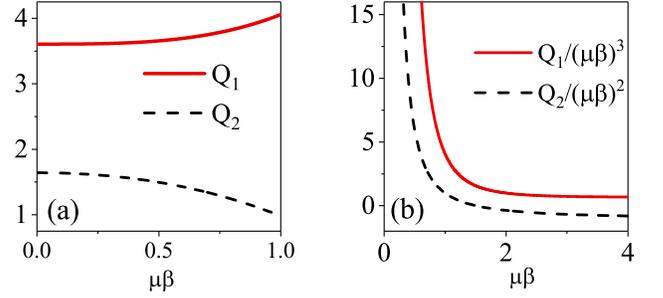}
\caption{The dimensionless function $Q_{1}$ and $Q_{2}$ in (a) and $Q_{1}/(\mu\beta)^{3}$ and $Q_{2}/(\mu\beta)^{3}$ in (b) are calculated with $\beta \mu$, respectively. $\mu$ is the chemical potential and measured from the conductance-band minimum. $T$ is temperature, and $\beta=1/k_\text{B}T$.}
\label{Berry-mu}
\end{figure}

Fig. \ref{Berry-mu} shows the dependence of the dimensionless functions $Q_{1(2)}$ on the chemical potential $\mu$ and temperature $k_\text{B}T$.  When the chemical potential is much smaller than the temperature, namely $\mu\ll k_\text{B}T$, both dimensionless functions $Q_{1(2)}$ tend to be constant and independent on the chemical potential and temperature [Fig. \ref{Berry-mu}(a)]. As a result, the quantity $\Lambda_{x}^\text{T}$ determining the non-linear Nernst effect behaves as $\Lambda_{x}^\text{T}\propto T$. In the regime, $\mu\gg k_\text{B}T,$
we observe that $Q_{1}$ shows $(\mu/k_\text{B}T)^3$ and $Q_{2}$ shows $(\mu/k_\text{B}T)^{2}$-dependence [Fig. \ref{Berry-mu}(b)]. Hence, for a fixed chemical potential, the non-linear Nernst current [Eq.(\ref{j-2D1})(\ref{Lam_end})] exhibits a transition from $\textbf{j}^\text{nl}_\text{A}\sim T^{-2}$ temperature dependence in the low temperature regime to a stronger $\textbf{j}^\text{nl}_\text{A}\sim T$ dependence in the high temperature.

In summary, we study the non-linear Nernst effect as a  second order response of temperature gradient. We  have derived the non-linear Nernst current by using the semiclassical framework of electron dynamics. The non-linear Nernst current, induced by the combination of Berry curvature and the non-equilibrium carrier distribution, flows transverse to temperature gradient in the time-reversal symmetry system but with inversion symmetry deliberately broken. Applying the uniaxial strain along the zigzag direction ($x$-direction) to $1H$-TMDCs can break the underlying $C_{3v}$ symmetry and leave only the mirror symmetry $M_{x},$ giving rise to non-linear Nernst current. Remarkably, the non-linear Nernst current is insensitive to  the chemical potential  in the ``low doping" or ``high temperature" regime ($\mu \ll k_\text{B}T$). In contrast, in the ``high doping" or  ``low temperature" regime~($\mu \gg k_\text{B}T$), the current shows $\textbf{j}^\text{nl}_\text{A}\propto \mu^{3}$ chemical potential dependence. For a fixed chemical potential,  the non-linear Nernst current exhibits $\textbf{j}^\text{nl}_\text{A}\sim T^{-2}$ temperature dependence in the low temperature regime to $\textbf{j}^\text{nl}_\text{A}\sim T^{}$ in the high temperature. In this paper our discussion focus on low energy effective Hamiltonian model of strained TMDCs. In fact,  the asymmetry distributions on BC and velocity are also expected to generate finite $\Lambda^\text{T}_{x}$ in high energy regime.~\cite{You2018} Finally we wish to remark that the proposed non-linear Nernst effect should be observed in
other 2D and 3D materials subject to the proper symmetry constraints.


This work is supported by the Fundamental Research Funds for the Central Universities and the NSFC (Grant No. 11674317). J.-S. Y. thank the IFW excellence programme. T.L. acknowledges support from NSF/EFRI- 1741660. GS and ZGZ are supported in part by the National Key R\&D Program of China (Grant No. 2018FYA0305800), the Strategic Priority Research Program of CAS (Grant Nos. XDB28000000, XB-D07010100),  the NSFC (Grant No. 11834014), and Beijing Municipal Science and Technology Commission (Grant No. Z118100004218001).


\begin{thebibliography}{41}%
\makeatletter
\providecommand \@ifxundefined [1]{%
 \@ifx{#1\undefined}
}%
\providecommand \@ifnum [1]{%
 \ifnum #1\expandafter \@firstoftwo
 \else \expandafter \@secondoftwo
 \fi
}%
\providecommand \@ifx [1]{%
 \ifx #1\expandafter \@firstoftwo
 \else \expandafter \@secondoftwo
 \fi
}%
\providecommand \natexlab [1]{#1}%
\providecommand \enquote  [1]{``#1''}%
\providecommand \bibnamefont  [1]{#1}%
\providecommand \bibfnamefont [1]{#1}%
\providecommand \citenamefont [1]{#1}%
\providecommand \href@noop [0]{\@secondoftwo}%
\providecommand \href [0]{\begingroup \@sanitize@url \@href}%
\providecommand \@href[1]{\@@startlink{#1}\@@href}%
\providecommand \@@href[1]{\endgroup#1\@@endlink}%
\providecommand \@sanitize@url [0]{\catcode `\\12\catcode `\$12\catcode
  `\&12\catcode `\#12\catcode `\^12\catcode `\_12\catcode `\%12\relax}%
\providecommand \@@startlink[1]{}%
\providecommand \@@endlink[0]{}%
\providecommand \url  [0]{\begingroup\@sanitize@url \@url }%
\providecommand \@url [1]{\endgroup\@href {#1}{\urlprefix }}%
\providecommand \urlprefix  [0]{URL }%
\providecommand \Eprint [0]{\href }%
\providecommand \doibase [0]{http://dx.doi.org/}%
\providecommand \selectlanguage [0]{\@gobble}%
\providecommand \bibinfo  [0]{\@secondoftwo}%
\providecommand \bibfield  [0]{\@secondoftwo}%
\providecommand \translation [1]{[#1]}%
\providecommand \BibitemOpen [0]{}%
\providecommand \bibitemStop [0]{}%
\providecommand \bibitemNoStop [0]{.\EOS\space}%
\providecommand \EOS [0]{\spacefactor3000\relax}%
\providecommand \BibitemShut  [1]{\csname bibitem#1\endcsname}%
\let\auto@bib@innerbib\@empty
\bibitem [{\citenamefont {Xiao}\ \emph {et~al.}(2010)\citenamefont {Xiao},
  \citenamefont {Chang},\ and\ \citenamefont {Niu}}]{Xiao2010}%
  \BibitemOpen
  \bibfield  {author} {\bibinfo {author} {\bibfnamefont {D.}~\bibnamefont
  {Xiao}}, \bibinfo {author} {\bibfnamefont {M.-C.}\ \bibnamefont {Chang}}, \
  and\ \bibinfo {author} {\bibfnamefont {Q.}~\bibnamefont {Niu}},\ }\href
  {\doibase 10.1103/revmodphys.82.1959} {\bibfield  {journal} {\bibinfo
  {journal} {Reviews of Modern Physics}\ }\textbf {\bibinfo {volume} {82}},\
  \bibinfo {pages} {1959} (\bibinfo {year} {2010})}\BibitemShut {NoStop}%
\bibitem [{\citenamefont {Nagaosa}\ \emph {et~al.}(2010)\citenamefont
  {Nagaosa}, \citenamefont {Sinova}, \citenamefont {Onoda}, \citenamefont
  {MacDonald},\ and\ \citenamefont {Ong}}]{Nagaosa2010}%
  \BibitemOpen
  \bibfield  {author} {\bibinfo {author} {\bibfnamefont {N.}~\bibnamefont
  {Nagaosa}}, \bibinfo {author} {\bibfnamefont {J.}~\bibnamefont {Sinova}},
  \bibinfo {author} {\bibfnamefont {S.}~\bibnamefont {Onoda}}, \bibinfo
  {author} {\bibfnamefont {A.~H.}\ \bibnamefont {MacDonald}}, \ and\ \bibinfo
  {author} {\bibfnamefont {N.~P.}\ \bibnamefont {Ong}},\ }\href {\doibase
  10.1103/revmodphys.82.1539} {\bibfield  {journal} {\bibinfo  {journal}
  {Reviews of Modern Physics}\ }\textbf {\bibinfo {volume} {82}},\ \bibinfo
  {pages} {1539} (\bibinfo {year} {2010})}\BibitemShut {NoStop}%
\bibitem [{\citenamefont {Xiao}\ \emph {et~al.}(2006)\citenamefont {Xiao},
  \citenamefont {Yao}, \citenamefont {Fang},\ and\ \citenamefont
  {Niu}}]{Xiao2006}%
  \BibitemOpen
  \bibfield  {author} {\bibinfo {author} {\bibfnamefont {D.}~\bibnamefont
  {Xiao}}, \bibinfo {author} {\bibfnamefont {Y.}~\bibnamefont {Yao}}, \bibinfo
  {author} {\bibfnamefont {Z.}~\bibnamefont {Fang}}, \ and\ \bibinfo {author}
  {\bibfnamefont {Q.}~\bibnamefont {Niu}},\ }\href {\doibase
  10.1103/PhysRevLett.97.026603} {\bibfield  {journal} {\bibinfo  {journal}
  {Phys. Rev. Lett.}\ }\textbf {\bibinfo {volume} {97}},\ \bibinfo {pages}
  {026603} (\bibinfo {year} {2006})}\BibitemShut {NoStop}%
\bibitem [{\citenamefont {Zhang}\ \emph {et~al.}(2008)\citenamefont {Zhang},
  \citenamefont {Tewari}, \citenamefont {Yakovenko},\ and\ \citenamefont
  {Sarma}}]{Zhang2008}%
  \BibitemOpen
  \bibfield  {author} {\bibinfo {author} {\bibfnamefont {C.}~\bibnamefont
  {Zhang}}, \bibinfo {author} {\bibfnamefont {S.}~\bibnamefont {Tewari}},
  \bibinfo {author} {\bibfnamefont {V.~M.}\ \bibnamefont {Yakovenko}}, \ and\
  \bibinfo {author} {\bibfnamefont {S.~D.}\ \bibnamefont {Sarma}},\ }\href
  {\doibase 10.1103/PhysRevB.78.174508} {\bibfield  {journal} {\bibinfo
  {journal} {Phys. Rev. B}\ }\textbf {\bibinfo {volume} {78}},\ \bibinfo
  {pages} {174508} (\bibinfo {year} {2008})}\BibitemShut {NoStop}%
\bibitem [{\citenamefont {Zhang}\ \emph {et~al.}(2009)\citenamefont {Zhang},
  \citenamefont {Tewari},\ and\ \citenamefont {Sarma}}]{Zhang2009}%
  \BibitemOpen
  \bibfield  {author} {\bibinfo {author} {\bibfnamefont {C.}~\bibnamefont
  {Zhang}}, \bibinfo {author} {\bibfnamefont {S.}~\bibnamefont {Tewari}}, \
  and\ \bibinfo {author} {\bibfnamefont {S.~D.}\ \bibnamefont {Sarma}},\ }\href
  {\doibase 10.1103/PhysRevB.79.245424} {\bibfield  {journal} {\bibinfo
  {journal} {Phys. Rev. B}\ }\textbf {\bibinfo {volume} {79}},\ \bibinfo
  {pages} {245424} (\bibinfo {year} {2009})}\BibitemShut {NoStop}%
\bibitem [{\citenamefont {Zhu}\ and\ \citenamefont {Berakdar}(2013)}]{Zhu2013}%
  \BibitemOpen
  \bibfield  {author} {\bibinfo {author} {\bibfnamefont {Z.-G.}\ \bibnamefont
  {Zhu}}\ and\ \bibinfo {author} {\bibfnamefont {J.}~\bibnamefont {Berakdar}},\
  }\href {\doibase 10.1088/1367-2630/15/7/073028} {\bibfield  {journal}
  {\bibinfo  {journal} {New Journal of Physics}\ }\textbf {\bibinfo {volume}
  {15}},\ \bibinfo {pages} {073028} (\bibinfo {year} {2013})}\BibitemShut
  {NoStop}%
\bibitem [{\citenamefont {{Deyo}}\ \emph {et~al.}(2009)\citenamefont {{Deyo}},
  \citenamefont {{Golub}}, \citenamefont {{Ivchenko}},\ and\ \citenamefont
  {{Spivak}}}]{Deyo2009}%
  \BibitemOpen
  \bibfield  {author} {\bibinfo {author} {\bibfnamefont {E.}~\bibnamefont
  {{Deyo}}}, \bibinfo {author} {\bibfnamefont {L.~E.}\ \bibnamefont {{Golub}}},
  \bibinfo {author} {\bibfnamefont {E.~L.}\ \bibnamefont {{Ivchenko}}}, \ and\
  \bibinfo {author} {\bibfnamefont {B.}~\bibnamefont {{Spivak}}},\ }\href@noop
  {} {\bibfield  {journal} {\bibinfo  {journal} {arXiv:0904.1917}\ } (\bibinfo
  {year} {2009})}\BibitemShut {NoStop}%
\bibitem [{\citenamefont {Moore}\ and\ \citenamefont
  {Orenstein}(2010)}]{Moore2010}%
  \BibitemOpen
  \bibfield  {author} {\bibinfo {author} {\bibfnamefont {J.~E.}\ \bibnamefont
  {Moore}}\ and\ \bibinfo {author} {\bibfnamefont {J.}~\bibnamefont
  {Orenstein}},\ }\href {\doibase 10.1103/physrevlett.105.026805} {\bibfield
  {journal} {\bibinfo  {journal} {Physical Review Letters}\ }\textbf {\bibinfo
  {volume} {105}},\ \bibinfo {pages} {026805} (\bibinfo {year}
  {2010})}\BibitemShut {NoStop}%
\bibitem [{\citenamefont {Low}\ \emph {et~al.}(2015)\citenamefont {Low},
  \citenamefont {Jiang},\ and\ \citenamefont {Guinea}}]{Low2015}%
  \BibitemOpen
  \bibfield  {author} {\bibinfo {author} {\bibfnamefont {T.}~\bibnamefont
  {Low}}, \bibinfo {author} {\bibfnamefont {Y.}~\bibnamefont {Jiang}}, \ and\
  \bibinfo {author} {\bibfnamefont {F.}~\bibnamefont {Guinea}},\ }\href
  {\doibase 10.1103/physrevb.92.235447} {\bibfield  {journal} {\bibinfo
  {journal} {Physical Review B}\ }\textbf {\bibinfo {volume} {92}},\ \bibinfo
  {pages} {235447} (\bibinfo {year} {2015})}\BibitemShut {NoStop}%
\bibitem [{\citenamefont {Sodemann}\ and\ \citenamefont
  {Fu}(2015)}]{Sodemann2015}%
  \BibitemOpen
  \bibfield  {author} {\bibinfo {author} {\bibfnamefont {I.}~\bibnamefont
  {Sodemann}}\ and\ \bibinfo {author} {\bibfnamefont {L.}~\bibnamefont {Fu}},\
  }\href {\doibase 10.1103/physrevlett.115.216806} {\bibfield  {journal}
  {\bibinfo  {journal} {Physical Review Letters}\ }\textbf {\bibinfo {volume}
  {115}},\ \bibinfo {pages} {216806} (\bibinfo {year} {2015})}\BibitemShut
  {NoStop}%
\bibitem [{\citenamefont {Eginligil}\ \emph {et~al.}(2015)\citenamefont
  {Eginligil}, \citenamefont {Cao}, \citenamefont {Wang}, \citenamefont {Shen},
  \citenamefont {Cong}, \citenamefont {Shang}, \citenamefont {Soci},\ and\
  \citenamefont {Yu}}]{Eginligil2015}%
  \BibitemOpen
  \bibfield  {author} {\bibinfo {author} {\bibfnamefont {M.}~\bibnamefont
  {Eginligil}}, \bibinfo {author} {\bibfnamefont {B.}~\bibnamefont {Cao}},
  \bibinfo {author} {\bibfnamefont {Z.}~\bibnamefont {Wang}}, \bibinfo {author}
  {\bibfnamefont {X.}~\bibnamefont {Shen}}, \bibinfo {author} {\bibfnamefont
  {C.}~\bibnamefont {Cong}}, \bibinfo {author} {\bibfnamefont {J.}~\bibnamefont
  {Shang}}, \bibinfo {author} {\bibfnamefont {C.}~\bibnamefont {Soci}}, \ and\
  \bibinfo {author} {\bibfnamefont {T.}~\bibnamefont {Yu}},\ }\href {\doibase
  10.1038/ncomms8636} {\bibfield  {journal} {\bibinfo  {journal} {Nature
  Communications}\ }\textbf {\bibinfo {volume} {6}},\ \bibinfo {pages} {7636}
  (\bibinfo {year} {2015})}\BibitemShut {NoStop}%
\bibitem [{\citenamefont {Quereda}\ \emph {et~al.}(2018)\citenamefont
  {Quereda}, \citenamefont {Ghiasi}, \citenamefont {You}, \citenamefont
  {van~den Brink}, \citenamefont {van Wees},\ and\ \citenamefont {van~der
  Wal}}]{Quereda2018}%
  \BibitemOpen
  \bibfield  {author} {\bibinfo {author} {\bibfnamefont {J.}~\bibnamefont
  {Quereda}}, \bibinfo {author} {\bibfnamefont {T.~S.}\ \bibnamefont {Ghiasi}},
  \bibinfo {author} {\bibfnamefont {J.-S.}\ \bibnamefont {You}}, \bibinfo
  {author} {\bibfnamefont {J.}~\bibnamefont {van~den Brink}}, \bibinfo {author}
  {\bibfnamefont {B.~J.}\ \bibnamefont {van Wees}}, \ and\ \bibinfo {author}
  {\bibfnamefont {C.~H.}\ \bibnamefont {van~der Wal}},\ }\href@noop {}
  {\bibfield  {journal} {\bibinfo  {journal} {arXiv:1803.08289}\ } (\bibinfo
  {year} {2018})},\ \Eprint {http://arxiv.org/abs/1803.08289} {arXiv:1803.08289
  [cond-mat.mtrl-sci]} \BibitemShut {NoStop}%
\bibitem [{\citenamefont {Xu}\ \emph {et~al.}(2018)\citenamefont {Xu},
  \citenamefont {Ma}, \citenamefont {Shen}, \citenamefont {Fatemi},
  \citenamefont {Wu}, \citenamefont {Chang}, \citenamefont {Chang},
  \citenamefont {Valdivia}, \citenamefont {Chan}, \citenamefont {Gibson},
  \citenamefont {Zhou}, \citenamefont {Liu}, \citenamefont {Watanabe},
  \citenamefont {Taniguchi}, \citenamefont {Lin}, \citenamefont {Cava},
  \citenamefont {Fu}, \citenamefont {Gedik},\ and\ \citenamefont
  {Jarillo-Herrero}}]{Xu2018}%
  \BibitemOpen
  \bibfield  {author} {\bibinfo {author} {\bibfnamefont {S.-Y.}\ \bibnamefont
  {Xu}}, \bibinfo {author} {\bibfnamefont {Q.}~\bibnamefont {Ma}}, \bibinfo
  {author} {\bibfnamefont {H.}~\bibnamefont {Shen}}, \bibinfo {author}
  {\bibfnamefont {V.}~\bibnamefont {Fatemi}}, \bibinfo {author} {\bibfnamefont
  {S.}~\bibnamefont {Wu}}, \bibinfo {author} {\bibfnamefont {T.-R.}\
  \bibnamefont {Chang}}, \bibinfo {author} {\bibfnamefont {G.}~\bibnamefont
  {Chang}}, \bibinfo {author} {\bibfnamefont {A.~M.~M.}\ \bibnamefont
  {Valdivia}}, \bibinfo {author} {\bibfnamefont {C.-K.}\ \bibnamefont {Chan}},
  \bibinfo {author} {\bibfnamefont {Q.~D.}\ \bibnamefont {Gibson}}, \bibinfo
  {author} {\bibfnamefont {J.}~\bibnamefont {Zhou}}, \bibinfo {author}
  {\bibfnamefont {Z.}~\bibnamefont {Liu}}, \bibinfo {author} {\bibfnamefont
  {K.}~\bibnamefont {Watanabe}}, \bibinfo {author} {\bibfnamefont
  {T.}~\bibnamefont {Taniguchi}}, \bibinfo {author} {\bibfnamefont
  {H.}~\bibnamefont {Lin}}, \bibinfo {author} {\bibfnamefont {R.~J.}\
  \bibnamefont {Cava}}, \bibinfo {author} {\bibfnamefont {L.}~\bibnamefont
  {Fu}}, \bibinfo {author} {\bibfnamefont {N.}~\bibnamefont {Gedik}}, \ and\
  \bibinfo {author} {\bibfnamefont {P.}~\bibnamefont {Jarillo-Herrero}},\
  }\href {\doibase 10.1038/s41567-018-0189-6} {\bibfield  {journal} {\bibinfo
  {journal} {Nature Physics}\ } (\bibinfo {year} {2018}),\
  10.1038/s41567-018-0189-6}\BibitemShut {NoStop}%
\bibitem [{\citenamefont {Ma}\ \emph {et~al.}(2018)\citenamefont {Ma},
  \citenamefont {Xu}, \citenamefont {Shen}, \citenamefont {MacNeill},
  \citenamefont {Fatemi}, \citenamefont {Chang}, \citenamefont {Valdivia},
  \citenamefont {Wu}, \citenamefont {Du}, \citenamefont {Hsu}, \citenamefont
  {Fang}, \citenamefont {Gibson}, \citenamefont {Watanabe}, \citenamefont
  {Taniguchi}, \citenamefont {Cava}, \citenamefont {Kaxiras}, \citenamefont
  {Lu}, \citenamefont {Lin}, \citenamefont {Fu}, \citenamefont {Gedik},\ and\
  \citenamefont {Jarillo-Herrero}}]{Ma2018}%
  \BibitemOpen
  \bibfield  {author} {\bibinfo {author} {\bibfnamefont {Q.}~\bibnamefont
  {Ma}}, \bibinfo {author} {\bibfnamefont {S.-Y.}\ \bibnamefont {Xu}}, \bibinfo
  {author} {\bibfnamefont {H.}~\bibnamefont {Shen}}, \bibinfo {author}
  {\bibfnamefont {D.}~\bibnamefont {MacNeill}}, \bibinfo {author}
  {\bibfnamefont {V.}~\bibnamefont {Fatemi}}, \bibinfo {author} {\bibfnamefont
  {T.-R.}\ \bibnamefont {Chang}}, \bibinfo {author} {\bibfnamefont {A.~M.~M.}\
  \bibnamefont {Valdivia}}, \bibinfo {author} {\bibfnamefont {S.}~\bibnamefont
  {Wu}}, \bibinfo {author} {\bibfnamefont {Z.}~\bibnamefont {Du}}, \bibinfo
  {author} {\bibfnamefont {C.-H.}\ \bibnamefont {Hsu}}, \bibinfo {author}
  {\bibfnamefont {S.}~\bibnamefont {Fang}}, \bibinfo {author} {\bibfnamefont
  {Q.~D.}\ \bibnamefont {Gibson}}, \bibinfo {author} {\bibfnamefont
  {K.}~\bibnamefont {Watanabe}}, \bibinfo {author} {\bibfnamefont
  {T.}~\bibnamefont {Taniguchi}}, \bibinfo {author} {\bibfnamefont {R.~J.}\
  \bibnamefont {Cava}}, \bibinfo {author} {\bibfnamefont {E.}~\bibnamefont
  {Kaxiras}}, \bibinfo {author} {\bibfnamefont {H.-Z.}\ \bibnamefont {Lu}},
  \bibinfo {author} {\bibfnamefont {H.}~\bibnamefont {Lin}}, \bibinfo {author}
  {\bibfnamefont {L.}~\bibnamefont {Fu}}, \bibinfo {author} {\bibfnamefont
  {N.}~\bibnamefont {Gedik}}, \ and\ \bibinfo {author} {\bibfnamefont
  {P.}~\bibnamefont {Jarillo-Herrero}},\ }\href {\doibase
  10.1038/s41586-018-0807-6} {\bibfield  {journal} {\bibinfo  {journal}
  {Nature}\ }\textbf {\bibinfo {volume} {565}},\ \bibinfo {pages} {337}
  (\bibinfo {year} {2018})}\BibitemShut {NoStop}%
\bibitem [{\citenamefont {Kang}\ \emph {et~al.}(2018)\citenamefont {Kang},
  \citenamefont {Li}, \citenamefont {Sohn}, \citenamefont {Shan},\ and\
  \citenamefont {Mak}}]{Kang2018}%
  \BibitemOpen
  \bibfield  {author} {\bibinfo {author} {\bibfnamefont {K.}~\bibnamefont
  {Kang}}, \bibinfo {author} {\bibfnamefont {T.}~\bibnamefont {Li}}, \bibinfo
  {author} {\bibfnamefont {E.}~\bibnamefont {Sohn}}, \bibinfo {author}
  {\bibfnamefont {J.}~\bibnamefont {Shan}}, \ and\ \bibinfo {author}
  {\bibfnamefont {K.~F.}\ \bibnamefont {Mak}},\ }\href@noop {} {\bibfield
  {journal} {\bibinfo  {journal} {arXiv e-prints}\ ,\ \bibinfo {eid}
  {arXiv:1809.08744}} (\bibinfo {year} {2018})},\ \Eprint
  {http://arxiv.org/abs/1809.08744} {arXiv:1809.08744 [cond-mat.mes-hall]}
  \BibitemShut {NoStop}%
\bibitem [{\citenamefont {Nakai}\ and\ \citenamefont
  {Nagaosa}(2018)}]{Nakai2018}%
  \BibitemOpen
  \bibfield  {author} {\bibinfo {author} {\bibfnamefont {R.}~\bibnamefont
  {Nakai}}\ and\ \bibinfo {author} {\bibfnamefont {N.}~\bibnamefont
  {Nagaosa}},\ }\href@noop {} {\bibfield  {journal} {\bibinfo  {journal} {arXiv
  e-prints}\ ,\ \bibinfo {eid} {arXiv:1812.02372}} (\bibinfo {year} {2018})},\
  \Eprint {http://arxiv.org/abs/1812.02372} {arXiv:1812.02372
  [cond-mat.mes-hall]} \BibitemShut {NoStop}%
\bibitem [{\citenamefont {Nandy}\ and\ \citenamefont
  {Sodemann}(2019)}]{Nandy2019}%
  \BibitemOpen
  \bibfield  {author} {\bibinfo {author} {\bibfnamefont {S.}~\bibnamefont
  {Nandy}}\ and\ \bibinfo {author} {\bibfnamefont {I.}~\bibnamefont
  {Sodemann}},\ }\href@noop {} {\bibfield  {journal} {\bibinfo  {journal}
  {arXiv e-prints}\ ,\ \bibinfo {eid} {arXiv:1901.04467}} (\bibinfo {year}
  {2019})},\ \Eprint {http://arxiv.org/abs/1901.04467} {arXiv:1901.04467
  [cond-mat.mes-hall]} \BibitemShut {NoStop}%
\bibitem [{\citenamefont {Wang}\ \emph {et~al.}(2012)\citenamefont {Wang},
  \citenamefont {Kalantar-Zadeh}, \citenamefont {Kis}, \citenamefont
  {Coleman},\ and\ \citenamefont {Strano}}]{Wang2012}%
  \BibitemOpen
  \bibfield  {author} {\bibinfo {author} {\bibfnamefont {Q.~H.}\ \bibnamefont
  {Wang}}, \bibinfo {author} {\bibfnamefont {K.}~\bibnamefont
  {Kalantar-Zadeh}}, \bibinfo {author} {\bibfnamefont {A.}~\bibnamefont {Kis}},
  \bibinfo {author} {\bibfnamefont {J.~N.}\ \bibnamefont {Coleman}}, \ and\
  \bibinfo {author} {\bibfnamefont {M.~S.}\ \bibnamefont {Strano}},\ }\href
  {\doibase 10.1038/nnano.2012.193} {\bibfield  {journal} {\bibinfo  {journal}
  {Nature Nanotechnology}\ }\textbf {\bibinfo {volume} {7}},\ \bibinfo {pages}
  {699} (\bibinfo {year} {2012})}\BibitemShut {NoStop}%
\bibitem [{\citenamefont {Xiao}\ \emph {et~al.}(2012)\citenamefont {Xiao},
  \citenamefont {Liu}, \citenamefont {Feng}, \citenamefont {Xu},\ and\
  \citenamefont {Yao}}]{Xiao2012}%
  \BibitemOpen
  \bibfield  {author} {\bibinfo {author} {\bibfnamefont {D.}~\bibnamefont
  {Xiao}}, \bibinfo {author} {\bibfnamefont {G.-B.}\ \bibnamefont {Liu}},
  \bibinfo {author} {\bibfnamefont {W.}~\bibnamefont {Feng}}, \bibinfo {author}
  {\bibfnamefont {X.}~\bibnamefont {Xu}}, \ and\ \bibinfo {author}
  {\bibfnamefont {W.}~\bibnamefont {Yao}},\ }\href {\doibase
  10.1103/physrevlett.108.196802} {\bibfield  {journal} {\bibinfo  {journal}
  {Physical Review Letters}\ }\textbf {\bibinfo {volume} {108}},\ \bibinfo
  {pages} {196802} (\bibinfo {year} {2012})}\BibitemShut {NoStop}%
\bibitem [{\citenamefont {Mak}\ \emph {et~al.}(2014)\citenamefont {Mak},
  \citenamefont {McGill}, \citenamefont {Park},\ and\ \citenamefont
  {McEuen}}]{Mak2014}%
  \BibitemOpen
  \bibfield  {author} {\bibinfo {author} {\bibfnamefont {K.~F.}\ \bibnamefont
  {Mak}}, \bibinfo {author} {\bibfnamefont {K.~L.}\ \bibnamefont {McGill}},
  \bibinfo {author} {\bibfnamefont {J.}~\bibnamefont {Park}}, \ and\ \bibinfo
  {author} {\bibfnamefont {P.~L.}\ \bibnamefont {McEuen}},\ }\href {\doibase
  10.1126/science.1250140} {\bibfield  {journal} {\bibinfo  {journal}
  {Science}\ }\textbf {\bibinfo {volume} {344}},\ \bibinfo {pages} {1489}
  (\bibinfo {year} {2014})}\BibitemShut {NoStop}%
\bibitem [{\citenamefont {Yu}\ \emph {et~al.}(2015)\citenamefont {Yu},
  \citenamefont {Zhu}, \citenamefont {Su},\ and\ \citenamefont
  {Jauho}}]{Yu2015}%
  \BibitemOpen
  \bibfield  {author} {\bibinfo {author} {\bibfnamefont {X.-Q.}\ \bibnamefont
  {Yu}}, \bibinfo {author} {\bibfnamefont {Z.-G.}\ \bibnamefont {Zhu}},
  \bibinfo {author} {\bibfnamefont {G.}~\bibnamefont {Su}}, \ and\ \bibinfo
  {author} {\bibfnamefont {A.-P.}\ \bibnamefont {Jauho}},\ }\href {\doibase
  10.1103/PhysRevLett.115.246601} {\bibfield  {journal} {\bibinfo  {journal}
  {Phys. Rev. Lett.}\ }\textbf {\bibinfo {volume} {115}},\ \bibinfo {pages}
  {246601} (\bibinfo {year} {2015})}\BibitemShut {NoStop}%
\bibitem [{\citenamefont {Yu}\ \emph {et~al.}(2017)\citenamefont {Yu},
  \citenamefont {Zhu}, \citenamefont {Su},\ and\ \citenamefont
  {Jauho}}]{Yu2017}%
  \BibitemOpen
  \bibfield  {author} {\bibinfo {author} {\bibfnamefont {X.-Q.}\ \bibnamefont
  {Yu}}, \bibinfo {author} {\bibfnamefont {Z.-G.}\ \bibnamefont {Zhu}},
  \bibinfo {author} {\bibfnamefont {G.}~\bibnamefont {Su}}, \ and\ \bibinfo
  {author} {\bibfnamefont {A.-P.}\ \bibnamefont {Jauho}},\ }\href {\doibase
  10.1103/PhysRevApplied.8.054038} {\bibfield  {journal} {\bibinfo  {journal}
  {Phys. Rev. Applied}\ }\textbf {\bibinfo {volume} {8}},\ \bibinfo {pages}
  {054038} (\bibinfo {year} {2017})}\BibitemShut {NoStop}%
\bibitem [{\citenamefont {You}\ \emph {et~al.}(2018)\citenamefont {You},
  \citenamefont {Fang}, \citenamefont {Xu}, \citenamefont {Kaxiras},\ and\
  \citenamefont {Low}}]{You2018}%
  \BibitemOpen
  \bibfield  {author} {\bibinfo {author} {\bibfnamefont {J.-S.}\ \bibnamefont
  {You}}, \bibinfo {author} {\bibfnamefont {S.}~\bibnamefont {Fang}}, \bibinfo
  {author} {\bibfnamefont {S.-Y.}\ \bibnamefont {Xu}}, \bibinfo {author}
  {\bibfnamefont {E.}~\bibnamefont {Kaxiras}}, \ and\ \bibinfo {author}
  {\bibfnamefont {T.}~\bibnamefont {Low}},\ }\href {\doibase
  10.1103/PhysRevB.98.121109} {\bibfield  {journal} {\bibinfo  {journal} {Phys.
  Rev. B}\ }\textbf {\bibinfo {volume} {98}},\ \bibinfo {pages} {121109 (R)}
  (\bibinfo {year} {2018})}\BibitemShut {NoStop}%
\bibitem [{\citenamefont {Lee}\ \emph {et~al.}(2017)\citenamefont {Lee},
  \citenamefont {Wang}, \citenamefont {Xie}, \citenamefont {Mak},\ and\
  \citenamefont {Shan}}]{Lee2017}%
  \BibitemOpen
  \bibfield  {author} {\bibinfo {author} {\bibfnamefont {J.}~\bibnamefont
  {Lee}}, \bibinfo {author} {\bibfnamefont {Z.}~\bibnamefont {Wang}}, \bibinfo
  {author} {\bibfnamefont {H.}~\bibnamefont {Xie}}, \bibinfo {author}
  {\bibfnamefont {K.~F.}\ \bibnamefont {Mak}}, \ and\ \bibinfo {author}
  {\bibfnamefont {J.}~\bibnamefont {Shan}},\ }\href {\doibase 10.1038/nmat4931}
  {\bibfield  {journal} {\bibinfo  {journal} {Nature Materials}\ }\textbf
  {\bibinfo {volume} {16}},\ \bibinfo {pages} {887} (\bibinfo {year}
  {2017})}\BibitemShut {NoStop}%
\bibitem [{\citenamefont {Zhang}\ \emph {et~al.}(2013)\citenamefont {Zhang},
  \citenamefont {Cheng}, \citenamefont {Gan},\ and\ \citenamefont
  {Schwingenschl\"{o}gl}}]{QingyuZhang2013}%
  \BibitemOpen
  \bibfield  {author} {\bibinfo {author} {\bibfnamefont {Q.}~\bibnamefont
  {Zhang}}, \bibinfo {author} {\bibfnamefont {Y.}~\bibnamefont {Cheng}},
  \bibinfo {author} {\bibfnamefont {L.-Y.}\ \bibnamefont {Gan}}, \ and\
  \bibinfo {author} {\bibfnamefont {U.}~\bibnamefont {Schwingenschl\"{o}gl}},\
  }\href {\doibase 10.1103/PhysRevB.88.245447} {\bibfield  {journal} {\bibinfo
  {journal} {Phys. Rev. B}\ }\textbf {\bibinfo {volume} {88}},\ \bibinfo
  {pages} {245447} (\bibinfo {year} {2013})}\BibitemShut {NoStop}%
\bibitem [{\citenamefont {Rodriguez-Lopez}\ and\ \citenamefont
  {Cortijo}(2018)}]{RodriguezLopez2018}%
  \BibitemOpen
  \bibfield  {author} {\bibinfo {author} {\bibfnamefont {P.}~\bibnamefont
  {Rodriguez-Lopez}}\ and\ \bibinfo {author} {\bibfnamefont {A.}~\bibnamefont
  {Cortijo}},\ }\href {\doibase 10.1103/PhysRevB.97.235128} {\bibfield
  {journal} {\bibinfo  {journal} {Phys. Rev. B}\ }\textbf {\bibinfo {volume}
  {97}},\ \bibinfo {pages} {235128} (\bibinfo {year} {2018})}\BibitemShut
  {NoStop}%
\bibitem [{\citenamefont {Fang}\ \emph {et~al.}(2018)\citenamefont {Fang},
  \citenamefont {Carr}, \citenamefont {Cazalilla},\ and\ \citenamefont
  {Kaxiras}}]{Fang2018}%
  \BibitemOpen
  \bibfield  {author} {\bibinfo {author} {\bibfnamefont {S.}~\bibnamefont
  {Fang}}, \bibinfo {author} {\bibfnamefont {S.}~\bibnamefont {Carr}}, \bibinfo
  {author} {\bibfnamefont {M.~A.}\ \bibnamefont {Cazalilla}}, \ and\ \bibinfo
  {author} {\bibfnamefont {E.}~\bibnamefont {Kaxiras}},\ }\href {\doibase
  10.1103/PhysRevB.98.075106} {\bibfield  {journal} {\bibinfo  {journal} {Phys.
  Rev. B}\ }\textbf {\bibinfo {volume} {98}},\ \bibinfo {pages} {075106}
  (\bibinfo {year} {2018})}\BibitemShut {NoStop}%
\bibitem [{\citenamefont {Gao}\ and\ \citenamefont {Xiao}(2018)}]{Gao2018}%
  \BibitemOpen
  \bibfield  {author} {\bibinfo {author} {\bibfnamefont {Y.}~\bibnamefont
  {Gao}}\ and\ \bibinfo {author} {\bibfnamefont {D.}~\bibnamefont {Xiao}},\
  }\href {\doibase 10.1103/PhysRevB.98.060402} {\bibfield  {journal} {\bibinfo
  {journal} {Phys. Rev. B}\ }\textbf {\bibinfo {volume} {98}},\ \bibinfo
  {pages} {060402 (R)} (\bibinfo {year} {2018})}\BibitemShut {NoStop}%
\bibitem [{\citenamefont {Morimoto}\ and\ \citenamefont
  {Nagaosa}(2016)}]{Morimoto2016}%
  \BibitemOpen
  \bibfield  {author} {\bibinfo {author} {\bibfnamefont {T.}~\bibnamefont
  {Morimoto}}\ and\ \bibinfo {author} {\bibfnamefont {N.}~\bibnamefont
  {Nagaosa}},\ }\href {\doibase 10.1103/PhysRevLett.117.146603} {\bibfield
  {journal} {\bibinfo  {journal} {Phys. Rev. Lett.}\ }\textbf {\bibinfo
  {volume} {117}},\ \bibinfo {pages} {146603} (\bibinfo {year}
  {2016})}\BibitemShut {NoStop}%
\bibitem [{\citenamefont {de~Juan}\ \emph {et~al.}(2017)\citenamefont
  {de~Juan}, \citenamefont {Grushin}, \citenamefont {Morimoto},\ and\
  \citenamefont {Moore}}]{deJuan2017}%
  \BibitemOpen
  \bibfield  {author} {\bibinfo {author} {\bibfnamefont {F.}~\bibnamefont
  {de~Juan}}, \bibinfo {author} {\bibfnamefont {A.~G.}\ \bibnamefont
  {Grushin}}, \bibinfo {author} {\bibfnamefont {T.}~\bibnamefont {Morimoto}}, \
  and\ \bibinfo {author} {\bibfnamefont {J.~E.}\ \bibnamefont {Moore}},\ }\href
  {\doibase 10.1038/ncomms15995} {\bibfield  {journal} {\bibinfo  {journal}
  {Nature Communications}\ }\textbf {\bibinfo {volume} {8}},\ \bibinfo {pages}
  {15995} (\bibinfo {year} {2017})}\BibitemShut {NoStop}%
\bibitem [{\citenamefont {Tsirkin}\ \emph {et~al.}(2018)\citenamefont
  {Tsirkin}, \citenamefont {Puente},\ and\ \citenamefont
  {Souza}}]{Tsirkin2018}%
  \BibitemOpen
  \bibfield  {author} {\bibinfo {author} {\bibfnamefont {S.~S.}\ \bibnamefont
  {Tsirkin}}, \bibinfo {author} {\bibfnamefont {P.~A.}\ \bibnamefont {Puente}},
  \ and\ \bibinfo {author} {\bibfnamefont {I.}~\bibnamefont {Souza}},\ }\href
  {\doibase 10.1103/physrevb.97.035158} {\bibfield  {journal} {\bibinfo
  {journal} {Physical Review B}\ }\textbf {\bibinfo {volume} {97}},\ \bibinfo
  {pages} {035158} (\bibinfo {year} {2018})}\BibitemShut {NoStop}%
\bibitem [{\citenamefont {Zhang}\ \emph
  {et~al.}(2018{\natexlab{a}})\citenamefont {Zhang}, \citenamefont {Sun},\ and\
  \citenamefont {Yan}}]{Zhang2018}%
  \BibitemOpen
  \bibfield  {author} {\bibinfo {author} {\bibfnamefont {Y.}~\bibnamefont
  {Zhang}}, \bibinfo {author} {\bibfnamefont {Y.}~\bibnamefont {Sun}}, \ and\
  \bibinfo {author} {\bibfnamefont {B.}~\bibnamefont {Yan}},\ }\href {\doibase
  10.1103/physrevb.97.041101} {\bibfield  {journal} {\bibinfo  {journal}
  {Physical Review B}\ }\textbf {\bibinfo {volume} {97}},\ \bibinfo {pages}
  {041101 (R)} (\bibinfo {year} {2018}{\natexlab{a}})}\BibitemShut {NoStop}%
\bibitem [{\citenamefont {Zhang}\ \emph
  {et~al.}(2018{\natexlab{b}})\citenamefont {Zhang}, \citenamefont {van~den
  Brink}, \citenamefont {Felser},\ and\ \citenamefont {Yan}}]{Zhang2018_1}%
  \BibitemOpen
  \bibfield  {author} {\bibinfo {author} {\bibfnamefont {Y.}~\bibnamefont
  {Zhang}}, \bibinfo {author} {\bibfnamefont {J.}~\bibnamefont {van~den
  Brink}}, \bibinfo {author} {\bibfnamefont {C.}~\bibnamefont {Felser}}, \ and\
  \bibinfo {author} {\bibfnamefont {B.}~\bibnamefont {Yan}},\ }\href {\doibase
  10.1088/2053-1583/aad1ae} {\bibfield  {journal} {\bibinfo  {journal} {2D
  Materials}\ }\textbf {\bibinfo {volume} {5}},\ \bibinfo {pages} {044001}
  (\bibinfo {year} {2018}{\natexlab{b}})}\BibitemShut {NoStop}%
\bibitem [{\citenamefont {Facio}\ \emph {et~al.}(2018)\citenamefont {Facio},
  \citenamefont {Efremov}, \citenamefont {Koepernik}, \citenamefont {You},
  \citenamefont {Sodemann},\ and\ \citenamefont {van~den Brink}}]{Facio2018}%
  \BibitemOpen
  \bibfield  {author} {\bibinfo {author} {\bibfnamefont {J.~I.}\ \bibnamefont
  {Facio}}, \bibinfo {author} {\bibfnamefont {D.}~\bibnamefont {Efremov}},
  \bibinfo {author} {\bibfnamefont {K.}~\bibnamefont {Koepernik}}, \bibinfo
  {author} {\bibfnamefont {J.-S.}\ \bibnamefont {You}}, \bibinfo {author}
  {\bibfnamefont {I.}~\bibnamefont {Sodemann}}, \ and\ \bibinfo {author}
  {\bibfnamefont {J.}~\bibnamefont {van~den Brink}},\ }\href {\doibase
  10.1103/PhysRevLett.121.246403} {\bibfield  {journal} {\bibinfo  {journal}
  {Phys. Rev. Lett.}\ }\textbf {\bibinfo {volume} {121}},\ \bibinfo {pages}
  {246403} (\bibinfo {year} {2018})}\BibitemShut {NoStop}%
\bibitem [{\citenamefont {Du}\ \emph {et~al.}(2018)\citenamefont {Du},
  \citenamefont {Wang}, \citenamefont {Lu},\ and\ \citenamefont
  {Xie}}]{Du2018}%
  \BibitemOpen
  \bibfield  {author} {\bibinfo {author} {\bibfnamefont {Z.~Z.}\ \bibnamefont
  {Du}}, \bibinfo {author} {\bibfnamefont {C.~M.}\ \bibnamefont {Wang}},
  \bibinfo {author} {\bibfnamefont {H.-Z.}\ \bibnamefont {Lu}}, \ and\ \bibinfo
  {author} {\bibfnamefont {X.~C.}\ \bibnamefont {Xie}},\ }\href {\doibase
  10.1103/PhysRevLett.121.266601} {\bibfield  {journal} {\bibinfo  {journal}
  {Phys. Rev. Lett.}\ }\textbf {\bibinfo {volume} {121}},\ \bibinfo {pages}
  {266601} (\bibinfo {year} {2018})}\BibitemShut {NoStop}%
\bibitem [{\citenamefont {{Du}}\ \emph {et~al.}(2018)\citenamefont {{Du}},
  \citenamefont {{Wang}}, \citenamefont {{Lu}},\ and\ \citenamefont
  {{Xie}}}]{Du2018_1}%
  \BibitemOpen
  \bibfield  {author} {\bibinfo {author} {\bibfnamefont {Z.~Z.}\ \bibnamefont
  {{Du}}}, \bibinfo {author} {\bibfnamefont {C.~M.}\ \bibnamefont {{Wang}}},
  \bibinfo {author} {\bibfnamefont {H.-Z.}\ \bibnamefont {{Lu}}}, \ and\
  \bibinfo {author} {\bibfnamefont {X.~C.}\ \bibnamefont {{Xie}}},\ }\href@noop
  {} {\bibfield  {journal} {\bibinfo  {journal} {arXiv e-prints}\ ,\ \bibinfo
  {eid} {arXiv:1812.08377}} (\bibinfo {year} {2018})},\ \Eprint
  {http://arxiv.org/abs/1812.08377} {arXiv:1812.08377 [cond-mat.mes-hall]}
  \BibitemShut {NoStop}%
\bibitem [{\citenamefont {kun Shi}\ and\ \citenamefont {Song}(2019)}]{Shi2019}%
  \BibitemOpen
  \bibfield  {author} {\bibinfo {author} {\bibfnamefont {L.}~\bibnamefont {kun
  Shi}}\ and\ \bibinfo {author} {\bibfnamefont {J.~C.~W.}\ \bibnamefont
  {Song}},\ }\href {\doibase 10.1103/PhysRevB.99.035403} {\bibfield  {journal}
  {\bibinfo  {journal} {Phys. Rev. B}\ }\textbf {\bibinfo {volume} {99}},\
  \bibinfo {pages} {035403 (R)} (\bibinfo {year} {2019})}\BibitemShut {NoStop}%
{\color{red}{\bibitem{SI} See Supplemental Material at [URL will be inserted by publisher] for the derviation of equations in main text, which includes Refs. \onlinecite{SI-ziman,SI-Landau,SI-Diao-1,SI-Cooper}}}
{\color{red}{\bibitem{SI-ziman} J. M. Ziman, \textit{Electrons and Phonons: the Theory of transport phenomena in solid}~(Oxford University Press, New York, 1960)}}
{\color{red}{\bibitem {SI-Landau} L. D. Landau and E. M. Lifshitz, \textit{Course of Theoretical Physics: statistical physics}~(Pergramon, Oxford, 1980)}}
{\color{red}{\bibitem{SI-Diao-1} D. Xiao, J. Shi, and Q. Niu, Phys. Rev. Lett. \textbf{95}, 137204~(2005)}}
{\color{red}{\bibitem{SI-Cooper} N. R. Cooper, B. I. Halperin, and I. M. Ruzin, Phys. Rev. B \textbf{55}, 2344~(1997)}}
\bibitem [{\citenamefont {Mahan}(2000)}]{mahan_many_2000}%
  \BibitemOpen
  \bibfield  {author} {\bibinfo {author} {\bibfnamefont {G.~D.}\ \bibnamefont
  {Mahan}},\ }\href@noop {} {\emph {\bibinfo {title} {Many Particle
  Physics}}},\ \bibinfo {edition} {3rd}\ ed.\ (\bibinfo  {publisher} {Kluwer},\
  \bibinfo {address} {New York},\ \bibinfo {year} {2000})\BibitemShut {NoStop}%
\bibitem [{\citenamefont {Rostami}\ \emph {et~al.}(2015)\citenamefont
  {Rostami}, \citenamefont {Rold{\'{a}}n}, \citenamefont {Cappelluti},
  \citenamefont {Asgari},\ and\ \citenamefont {Guinea}}]{Rostami2015}%
  \BibitemOpen
  \bibfield  {author} {\bibinfo {author} {\bibfnamefont {H.}~\bibnamefont
  {Rostami}}, \bibinfo {author} {\bibfnamefont {R.}~\bibnamefont
  {Rold{\'{a}}n}}, \bibinfo {author} {\bibfnamefont {E.}~\bibnamefont
  {Cappelluti}}, \bibinfo {author} {\bibfnamefont {R.}~\bibnamefont {Asgari}},
  \ and\ \bibinfo {author} {\bibfnamefont {F.}~\bibnamefont {Guinea}},\ }\href
  {\doibase 10.1103/PhysRevB.92.195402} {\bibfield  {journal} {\bibinfo
  {journal} {Phys. Rev. B}\ }\textbf {\bibinfo {volume} {92}},\ \bibinfo
  {pages} {195402} (\bibinfo {year} {2015})}\BibitemShut {NoStop}%
\bibitem [{\citenamefont {Liu}\ \emph {et~al.}(2013)\citenamefont {Liu},
  \citenamefont {Shan}, \citenamefont {Yao}, \citenamefont {Yao},\ and\
  \citenamefont {Xiao}}]{Liu2013}%
  \BibitemOpen
  \bibfield  {author} {\bibinfo {author} {\bibfnamefont {G.-B.}\ \bibnamefont
  {Liu}}, \bibinfo {author} {\bibfnamefont {W.-Y.}\ \bibnamefont {Shan}},
  \bibinfo {author} {\bibfnamefont {Y.}~\bibnamefont {Yao}}, \bibinfo {author}
  {\bibfnamefont {W.}~\bibnamefont {Yao}}, \ and\ \bibinfo {author}
  {\bibfnamefont {D.}~\bibnamefont {Xiao}},\ }\href {\doibase
  10.1103/PhysRevB.88.085433} {\bibfield  {journal} {\bibinfo  {journal} {Phys.
  Rev. B}\ }\textbf {\bibinfo {volume} {88}},\ \bibinfo {pages} {085433}
  (\bibinfo {year} {2013})}\BibitemShut {NoStop}%
\end{thebibliography}
\end{document}